\documentstyle[graphicx]{elsart}

\begin{document}
\begin{frontmatter}

\title{The $g_{V\sigma\gamma}$-coupling constants in hadron electrodynamics}
\author{Michail P. Rekalo \thanksref{addr}}
\thanks[addr]{ Permanent address:
\it National Science Center KFTI, 310108 Kharkov, Ukraine}
\address{Middle East Technical University,
Physics Department, Ankara 06531, Turkey}
\author{Egle Tomasi-Gustafsson\thanksref{corr}}
\thanks[corr]{Corresponing author: etomasi@cea.fr}
\address{\it DAPNIA/SPhN, CEA/Saclay, 91191 Gif-sur-Yvette Cedex,
France}

\begin{abstract}

Recent measurements of the branching ratios for the decays $\omega\to\pi^0\pi^0\gamma$ and $\rho^0\to\pi^0\pi^0\gamma$  lead to coupling constants of the  $V\sigma\gamma$-interaction ($V$ is a vector meson) one order of magnitude smaller than previously assumed to describe the threshold  cross section of $\gamma+p\to p+\rho^0$ and the HERMES effect. The new $g^2_{V\sigma\gamma}$ couplings are in contradiction with the predictions of the QCD sum rules, but are in good agreement with VDM-estimation of the $\sigma\to \gamma\gamma$-width.
\end{abstract}
\end{frontmatter}
\section{Introduction}
The coupling constants  
$g_{V\sigma\gamma}$, $V=\rho$ or $\omega$ are important ingredients for the theoretical analysis of many different hadronic electromagnetic processes. If $m_{\sigma}<m_V$, the decays $V\to \sigma+\gamma $ can occur directly-with radiation of electric dipole photons. In this respect, these decays are essentially different from the decays $V\to P+\gamma$, $P=\pi,$ $\eta$, $\eta '$ -with radiation of magnetic dipole photons. In the quark models, these last decays are induced by the quark magnetic moment, with transition $S=1\to S=0$, where $S$ is the total spin of the $q\overline{q}$-system (in the corresponding meson), whereas the decays $V\to\sigma+\gamma$ are induced by the internal motion of quarks, with transition $S=1\to S=1$ \cite{Ia92}.

Let us mention the main applications of the $g_{V\sigma\gamma}$-coupling constants:
\begin{itemize}
\item The $\sigma$-exchange for the vector meson production \cite{Fr96,Zh98,Ti99,La00,Ba02} $\gamma+p\to p+\rho^0$, near threshold.
\item The electromagnetic transition $\gamma^*\to \sigma+\omega$, in the space-like region of the virtual photon $\gamma^*$four-momenta, enters in the calculation of meson exchange currents, in particular in the analysis of the deuteron electromagnetic form factors at large momentum transfer \cite{Ch74,Hu90,VO95}.
\item The branching ratio of rare radiative decays of vector mesons as $\rho^0\to \pi^0\pi^0\gamma$ and $\omega\to \pi^0\pi^0\gamma$ are controlled by the $g_{V\sigma\gamma}$-coupling constants \cite{Re69,Go00}.

\item The HERMES effect \cite{Ac99}, concerning the inclusive electroproduction cross section on light nuclei can be explained under specific assumptions \cite{Mi00}  about the absolute value of the $g_{\rho\sigma\gamma}$ and $g_{\omega\sigma\gamma}$ coupling constants and the corresponding electromagnetic form factors.
\item The coupling constant of the $\sigma\to\gamma \gamma$-vertex, which can be calculated on the basis of the $g_{V\sigma\gamma}$ couplings, is important for the analysis of real and virtual Compton scattering on nucleons \cite{Ko98,Pa95,Gu79}.

\item The exact value of the $g_{V\sigma\gamma}$ coupling constant, in principle, allows one to constrain the $g_{\sigma NN}$-coupling, which describes the scalar exchange in the  $NN$-potential \cite{Ma89}.
\item  Finally, the  $g_{V\sigma\gamma}$ coupling constants have an implicit theoretical interest, being fundamental coupling constants of hadron electrodynamics.
\end{itemize}

Attempts to estimate these coupling constants from $QCD$-sum rules, give relatively large absolute values \cite{Go01}. Phenomenological methods based on the existing experimental data about different electromagnetic processes need serious additional assumptions. For example, in the framework of $\sigma$-exchange for $\gamma+p\to p+\rho^0$, it is possible to estimate the product $g_{\rho\sigma\gamma}g_{\sigma NN}$, assuming a definite form of the phenomenological form factors, which insure the correct behavior of the differential cross section $d\sigma/dt$, in the near-threshold region, and depend on ad-hoc cut-off parameters. Taking $g^2_{\sigma NN}/4\pi=8$, one obtains $g_{\rho\sigma\gamma}=2.71$ \cite{Fr96}. Considering only $\sigma$-exchange in $\gamma+p\to p+\rho^0$, the sign of this constant can not be determined, but some polarization observables are sensitive to this sign \cite{Re03}. Other possible contributions to the matrix element for the process $\gamma+p\to p+\rho^0$, such as $N^*$-excitation, $N$-exchange in $s$- and $u$-channel are neglected in such oversimplified consideration \cite{Fr96}.

Also the estimation of the discussed coupling constants done in framework of the HERMES effect \cite{Mi00} can not be considered as direct and model independent.

Another source of information about the $\rho \sigma\gamma$-vertex is the radiative decay $\rho^0\to\pi^+\pi^-\gamma$-with relatively large branching ratio \cite{PdG}. Namely this decay has been considered to favor large values of $g_{\rho\sigma\gamma}$ in comparison with $g_{\omega\sigma\gamma}$,  because of the corresponding widths $\Gamma(\omega\to\pi^+\pi^-\gamma)\ll \Gamma (\rho\to\pi^+\pi^-\gamma)$ \cite{Fr96}. However it is necessary to point out that for the decay $\rho^0\to\pi^+\pi^-\gamma$ the bremsstrahlung mechanism dominates, therefore the $\sigma$-contribution, being of the order of the error bars, can not give the $g_{\rho\sigma\gamma}$ with good accuracy. Note also that the precise determination of 
$g_{\rho\sigma\gamma}$  depends essentially on the mass and the width of the $\sigma$-meson. In \cite{Go00} different values of $g_{\rho\sigma\gamma}$ could been found, with the general conclusion that the  $g_{\rho\sigma\gamma}$ value from $\rho^0\to\pi^+\pi^-\gamma$ is not in contradiction with the estimation done in \cite{Fr96}. 

Finally, having a definite value of $g_{\rho\sigma\gamma}$,  it is possible to estimate the branching ratio for $\rho\to\pi^0\pi^0\gamma$, in framework of the effective Lagrangian approach. In this case, the absence of the bremsstrahlung mechanism results in the fact that the $\sigma$-exchange mechanism is important. But the predicted 
$BR(\rho^0\to \pi^0\pi^0\gamma)\simeq (45-200)\cdot 10^{-5}$,  with $g_{\rho\sigma\gamma}=3\div 5$, were too large in comparison with other theoretical expectations for this decay \cite{Fa90}. 

Moreover, and this will be the important background of this paper, large values of $g_{\rho\sigma\gamma}$ contradict essentially the recent experimental results from Novosibirsk, concerning the direct measurement of the decay $\rho^0\to \pi^0\pi^0\gamma$ \cite{Ac02}:
$$Br(\rho\to\pi^0\pi^0\gamma)=(4.1^{+1.0}_{-0.9}\pm 0.3)\cdot 10^{-5},$$
$$Br(\rho\to\sigma\gamma\to\pi^0\pi^0\gamma)=(1.9^{+0.9}_{- 0.8 }\pm 0.4)\cdot 10^{-5}.$$

The main aim of this paper is to estimate the coupling constant $g_{\rho\sigma\gamma}$ on the basis of these new important experimental data, which is, in our opinion, the most straightforward way.

\section{The $\sigma$-contribution to the $\rho\to\pi^0+\pi^0+\gamma$-decay}

A crude estimation of the $g_{\rho\sigma\gamma}$ coupling constant 
can be easily obtained from the experimental data about 
$\rho^0\to \sigma\gamma\to\pi^0\pi^0\gamma$ \cite{Ac02}, under the assumption that  the $\sigma$-mass is smaller than the $\rho$-meson mass, so that the decay $\rho\to\sigma+\gamma$ is allowed.

Neglecting for a moment the $\sigma$-width (which is an evident oversimplification of reality, as $\Gamma_{\sigma}=(0.6\div 1)$ GeV \cite{PdG}), we can find:
\begin{equation}
\Gamma(\rho^0\to\sigma\gamma\to\pi^0\pi^0\gamma)=\displaystyle\frac{1}{3}
\Gamma(\rho^0\to\sigma\gamma),
\label{eq:eq1}
\end{equation}
taking into account the identity of the neutral $\pi^0$-mesons, produced in 
$\rho^0\to\pi^0\pi^0\gamma$, and the isotopic relation: $g(\sigma\to\pi^0\pi^0)=g(\sigma\to\pi^+\pi^-)$.

The matrix element of the decay $V\to\sigma+\gamma$ can be written in the following form:
\begin{equation}
{\cal M}= \displaystyle\frac{eg_{V\sigma \gamma}}{M}\left(\epsilon^*\cdot U~k\cdot p -
\epsilon^*\cdot p~U\cdot k\right ),
\label{eq:eq2}
\end{equation}
where $\epsilon_{\mu}(k)$ and $U_{\mu}(p)$ are the four-vectors describing the vector polarization (four-momenta) of the photon and of the $V-$meson.

Averaging over the $V$-meson polarizations and summing over the photon polarizations, one can find for the width of the decay $V\to\sigma+\gamma$ (still considering the $\sigma$-meson as a stable particle):
\begin{equation}
\Gamma(V\to\sigma\gamma)=\alpha \displaystyle\frac{g^2_{V\sigma \gamma}}{24}M \left(1-\displaystyle\frac{m_{\sigma}^2}{M^2}\right )^3\simeq 231 g^2_{V\sigma \gamma}\left(1-\displaystyle\frac{m_{\sigma}^2}{M^2}\right )^3 \mbox{~keV},
\label{eq:eq3}
\end{equation}
where $M(m_{\sigma})$ is the mass of the $V(\sigma)$-meson. The numerical estimation (\ref{eq:eq3}) is done for M=0.77 GeV ($\rho$-meson). Evidently, at a given value of the $g_{V\sigma \gamma}$-coupling constant, the value of $\Gamma(V\to\sigma\gamma)$ depends essentially on the value taken for the $\sigma$-meson mass, which can be in a wide range: $m_{\sigma}=(0.4\div 1.2)$ GeV \cite{PdG}. In Table \ref{Table1} we report the values of 
$\Gamma(\rho^0\to\sigma\gamma)$ 
and $\Gamma(\rho^0\to\sigma\gamma\to\pi^0\pi^0\gamma)$, 
calculated for the values of the $g_{\rho\sigma \gamma}$-constant and $m_{\sigma}$ from Refs. \cite{Fr96} and \cite{Mi00}. Such estimations for $\Gamma(\rho^0\to\sigma\gamma\to\pi^0\pi^0\gamma)$ are almost two orders of magnitude larger than the value given by  the experiment \cite{Ac02}:
\begin{equation}
\Gamma(\rho^0\to\sigma\gamma\to\pi^0\pi^0\gamma)= (2.85^{+1.4}_{-1.2}\pm 0.6) ~\mbox{keV}.
\label{eq:eq5c}
\end{equation}

Note that the predictions of QCD sum rules \cite{Go01} are also in serious disagreement with  the direct experimental estimation (\ref{eq:eq5c}).

\begin{table}[h]
\begin{tabular}{|c|c|c|c|c|}
\hline\noalign{\smallskip}
$m_{\sigma}$&$g_{\rho\sigma \gamma}$&$ \Gamma(\rho^0\to\sigma\gamma)$ &$\Gamma(\rho^0\to\pi^0\pi^0\gamma)$&Ref. \\
$\mbox{[GeV]}$ & & $\mbox{[keV]}$ &$\mbox{[keV]}$ & \\
\noalign{\smallskip}\hline\noalign{\smallskip}
0.5&3& 379&126& \protect\cite{Fr96}\\
\noalign{\smallskip}\hline\noalign{\smallskip}
0.6&5-6& 308-444&103-148& \protect\cite{Mi00}\\
\noalign{\smallskip}\hline
\hline
\end{tabular}
\caption{Radiative widths for different masses of $\sigma$-meson, $m_{\sigma}$ and coupling constant $g_{\rho\sigma \gamma}$.}
\label{Table1}
\end{table}
This crude preliminary estimation leads to a conclusion  which is inconsistent with the suggested interpretation of the HERMES effect \cite{Mi00} and the description of threshold behavior of the cross section for $\gamma+p\to p+\rho^0$ \cite{Fr96}. Therefore let us estimate more precisely the $g_{\rho\sigma\gamma}$ coupling constant on the basis of the decay $\rho^0\to\sigma\gamma\to\pi^0\pi^0\gamma$, removing the hypothesis taken above, i.e. taking into account the finite value of the $\sigma$-width, and the possibility that in some cases it is possible to have $m_{\sigma}>M$.

\begin{figure}
\begin{center}
\includegraphics[height =10cm]{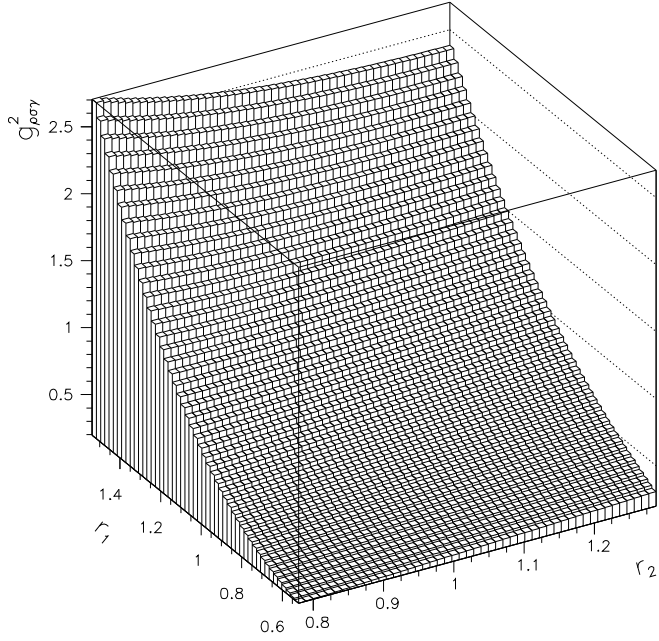}
\caption{Two-dimensional plot of the coupling constant $g^2_{\rho\sigma\gamma}$ as a function of $r_1=\displaystyle\frac{m_{\sigma}}{M}$ and $r_2=\displaystyle\frac{\Gamma_{\sigma}}{M}$. }
\label{fig:fig1}
\end{center}
\end{figure}

Considering the sequence of decays $\rho^0\to\sigma+\gamma\to\pi^0+\pi^0+\gamma$, one can find, for the effective $\pi^0+\pi^0$-mass distribution, the following formula (see Appendix):
\begin{equation}
\displaystyle\frac{d\Gamma}{dw^2}=\displaystyle\frac{e^2g^2_{\rho\sigma \gamma}}{288\pi^2}M\left(1-\displaystyle\frac{w^2}{M^2}\right )^3\displaystyle\frac{\beta_w}{\beta_{\sigma}}\displaystyle\frac{m_{\sigma}
\Gamma_{\sigma}}{(w^2-m_{\sigma}^2)^2+\Gamma^2_{\sigma} m_{\sigma}^2},
\label{eq:eq7}
\end{equation}
where $\Gamma_{\sigma}$ is the total width for the $\sigma$-meson, and:
$$\beta_{\sigma}=\sqrt{1- \displaystyle\frac{4m_{\pi}^2}{m_{\sigma}^2}},~ 
\beta_w=\sqrt{1- \displaystyle\frac{4m_{\pi}^2}{w^2}}.$$
For the estimation of the coupling constant $g^2_{\rho\sigma \gamma}$ on the basis of $\Gamma (\rho^0\to\sigma\gamma\to\pi^0\pi^0\gamma)$ the following formula can be used:
\begin{equation}
g^2_{\rho\sigma \gamma}= \displaystyle\frac{\Gamma(\rho^0\to\sigma\gamma\to\pi^0\pi^0\gamma)72\pi}
{M\alpha}
\displaystyle\frac{1}{r_1r_2 f(r_1,r_2)}
\label{eq:eq8}
\end{equation}
with
$$
f(r_1,r_2)= \int_a^1 dx\displaystyle\frac{(1-x)^3}{(x-r_1^2)^2+r_1^2r_2^2}
\left (\displaystyle\frac{1-a/x}{1-a/r_1^2}\right )^{1/2},~r_1=\displaystyle\frac{m_{\sigma}}{M},~r_2=\displaystyle\frac{\Gamma_{\sigma}}{M},
~a=\displaystyle\frac{4m_{\pi}^2}{M^2}.$$
Using the experimental information for the branching ratio for the radiative decay $\rho^0\to\sigma\gamma\to\pi^0+\pi^0+\gamma$ \cite{Ac02}, with the help of Eq. 
(\ref{eq:eq8}), one can deduce the constant 
$g_{\rho\sigma\gamma}$ as a function of the $\sigma$-meson parameters, its mass and its width (Fig. \ref{fig:fig1}). One can see a strong dependence of the value of $g_{\rho\sigma\gamma}$ on the two coordinates $r_1$ and $r_2$, which vary in the wide interval suggested by \cite{PdG}. Moreover, in the whole region of parameters $r_1$ and $r_2$ one can see that $g^2_{\rho\sigma\gamma}\le 2.5$; for example, for $m_{\sigma}=0.5$ GeV and $\Gamma_{\sigma}=0.6$ GeV, we find $g^2_{\rho\sigma\gamma}=0.196$.

\section{Estimation of $\lowercase{g}_{V\sigma\gamma}$ couplings from $\sigma\to 2\gamma$}
\begin{figure}
\begin{center}
\includegraphics[height =5cm]{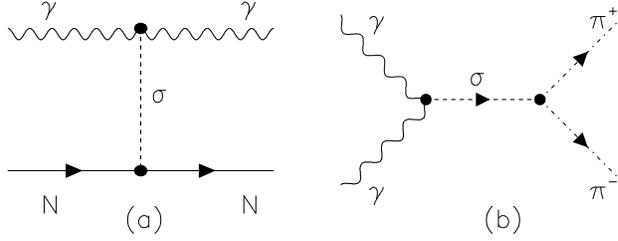}
\caption{Feynman diagrams for $\sigma$-meson contribution to nucleon Compton scattering (a) and to $\pi^+\pi^-$-production in $\gamma\gamma$-collisions (b).}
\label{fig:fig2}\end{center}
\end{figure}

Another source for an independent estimation of the $g_{\rho\sigma\gamma}$ coupling constant, 
which confirms our previous finding, is the radiative decay $\sigma\to 2\gamma$. In principle it is possible to estimate the width of this decay analyzing the $\sigma$-contribution in the amplitude of the elastic scattering of photons by nucleons, in the framework of the $t-$channel exchange, Fig. \ref{fig:fig2}a, or in the two-photon production of a $\pi^+\pi^-$ or $\pi^0\pi^0$-pair in $\gamma\gamma$-collisions, Fig. \ref{fig:fig2}b, in the corresponding region of the $\pi\pi$-effective mass \cite{Bo99}.

\begin{figure}
\begin{center}
\includegraphics[height =5cm]{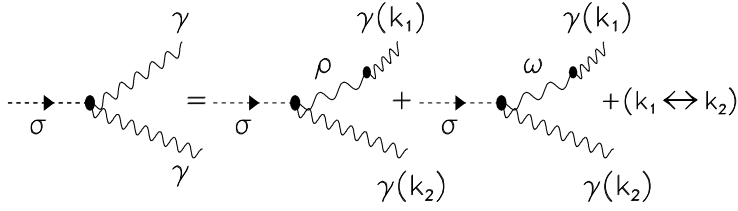}
\caption{Feynman diagrams for $\sigma\to 2\gamma$-decay in the VDM.}
\label{fig:fig3}\end{center}
\end{figure}

Following the VDM approach, (Fig. \ref{fig:fig3}), the matrix element for $\sigma\to 2\gamma$ can be written in terms of the $g_{V\sigma\gamma}$ coupling constants, as follows:
\begin{equation}
{\cal M}(\sigma\to 2\gamma)=e^2(g_{\rho}g_{\rho\sigma\gamma}+g_{\omega}g_{\omega\sigma\gamma})\vec e_1\cdot \vec e_2 m_{\sigma}^2/M,
\label{eq:eq9}
\end{equation}
where $\vec e_1$ and $\vec e_2$ are the three-vectors of polarization for the two produced photons. The coupling constant $g_V$ determines the width of the  $V$-meson leptonic decay 
$V\to e^+e^-$ (through one-photon exchange): $\Gamma(V\to e^+e^-)=\alpha^2g_V^2M\displaystyle\frac{4\pi}{3}$. Note that this formula holds for zero lepton mass. The existing data about 
$\rho(\omega)\to e^+e^-$ decays \cite{PdG} allow to estimate $g_{\rho}$ and $g_{\omega}$ with good accuracy:
$g_{\rho}^2\simeq 0.04,~~g_{\omega}^2\simeq 0.09 g_{\rho}^2$.

As both couplings $g_{\rho\sigma\gamma}$ and $g_{\omega\sigma\gamma}$ enter in Eq. (\ref{eq:eq9}), we assume, for simplicity, the validity of the $SU(3)$ relation:  $g_{\rho\sigma\gamma}/g_{\omega\sigma\gamma}\simeq 3$. The width $\Gamma(\sigma\to 2\gamma)$ can be written as: 
\begin{equation}
\Gamma(\sigma\to 2\gamma)=\pi\alpha^2
\displaystyle\frac{m_{\sigma}^3}{M^2}g^2_{\rho}g^2_{\rho\sigma\gamma}
\left (1+\displaystyle\frac{g_{\omega}}
{3g_{\rho}}\right )^2 \simeq  13.7 g^2_{\rho\sigma \gamma}
\left (\displaystyle\frac{m_{\sigma}}{1 \mbox{~GeV}}\right )^3 \mbox{~keV}.
\label{eq:eq12}
\end{equation}
Taking $g_{\rho\sigma \gamma}\simeq 3$ \cite{Fr96}, one finds 
\begin{equation}
\Gamma(\sigma\to 2\gamma)=123\mbox{~keV},~ \mbox{for} ~m_{\sigma}=1~ \mbox{GeV}.
\label{eq:eq13}
\end{equation}
The same or even larger numbers for 
$\Gamma(\sigma\to 2\gamma)$ follow from the QCD-sum rule estimation for $g_{\rho\sigma\gamma}$ \cite{Go01} or from the value deduced from the analysis of $\rho^0\to\pi^+\pi^-\gamma$-decays \cite{Go00}. These numbers contradict the corresponding experimental estimations: 
\begin{equation}
\Gamma(\sigma\to 2\gamma)=10.6 \mbox{~keV~}  \protect\cite{PdG},~  \Gamma(\sigma\to 2\gamma)=(3.8\pm 1.5) \mbox{~keV~} \protect\cite{Bo99}.
\label{eq:eq14}
\end{equation}
And, again, the discrepancy is large, up to one order magnitude. The situation is even worse, taking the value of $g_{\rho\sigma \gamma}$ from Ref. \cite{Mi00}, which leads to 
\begin{equation} 
\Gamma(\sigma\to 2\gamma)=(340 \div 490) \mbox{~keV}.
\label{eq:eq15}
\end{equation}

On the other hand, taking the experimental data about the $\sigma\to 2\gamma$-decay, (\ref{eq:eq14}), one can deduce directly a value 
\begin{equation} 
0.28 \le g^2_{\rho\sigma \gamma}
\left (\displaystyle\frac{m_{\sigma}}{1 \mbox{~GeV}}\right )^3\le 0.78,
\label{eq:eq16}
\end{equation}
in evident contradiction with the previous large 
$g_{\rho\sigma \gamma}$ values \cite{Fr96,Go00,Mi00,Go01}, but in good agreement with the present estimation, based on the new data about the decay $\rho\to\pi^0\pi^0\gamma$ \cite{Ac02}.

\section{Discussion and conclusions}
Let us discuss the possible consequences of a smaller value of the $g_{\rho\sigma\gamma}$ coupling constant; more precisely, let us consider the case $g^2_{\rho\sigma\gamma}\le 1$. Here this range is derived   from two independent sources: the radiative decay of the $\rho^0$-meson, $\rho^0\to\sigma\gamma\to\pi^0+\pi^0+\gamma$, on one side, and the radiative decay of the $\sigma$-meson:  
$\sigma\to 2\gamma$, from another side. Both these decays can be described in a relative simple and transparent theoretical framework: the  $\sigma$-dominance for the first one and the VDM approach for 
the second one. Note that the VDM model gives a good description of the different numerous radiative decays involving vector and pseudoscalar mesons. The coupling constants $g_{\rho}$ and $g_{\omega}$ (for the $\gamma\to V^0$-transition), which 
enter in our consideration of the decay $\sigma\to 2\gamma$, are well known from the experimental data about the decays $V\to\ell^+\ell^-$, which have quite good accuracy. Therefore, the main uncertainty in the determination of $g_{\rho\sigma\gamma}$ on the basis of the existing data on $\sigma\to 2\gamma$, derives from the relatively large interval for 
$\Gamma( \sigma\to 2\gamma)$. The parameters of the $\sigma$-meson also do not affect very much this estimation: there is no dependence on the  $\sigma$-width, and only a cubic dependence on the $\sigma-$meson mass, so that:
\begin{equation}
g^2_{\rho\sigma\gamma}=
(g^2_{\rho\sigma\gamma})_0\left (\displaystyle\frac{m_{\sigma}}{1~\mbox{GeV}}\right )^{-3},
\label{eq:eq17}
\end{equation}
where$(g^2_{\rho\sigma\gamma})_0$ is the value of the considered coupling constant for $m_{\sigma}$=1 GeV. From this scaling law it follows the $g_{\rho\sigma\gamma}$ constant  gets smaller  for higher $\sigma$-masses. 

The estimation of $g^2_{\rho\sigma\gamma}$ on the basis of the decay $\rho^0\to\sigma+\gamma\to\pi^0+\pi^0+\gamma$ depends essentially on the mass and width of the $\sigma$-meson, only, without additional unknown coupling constants. We can build a  two-dimensional representation on $g^2_{\rho\sigma\gamma}$, with non trivial dependence on $m_{\sigma} $ and $\Gamma_{\sigma} $. It turns out  that, for the 'standard' value  $m_{\sigma} $ =0.5 GeV, in the interval $0.6\le\Gamma_{\sigma}\le 1 $ GeV, one has the following limits:
\begin{equation}
0.19\le g^2_{\rho\sigma\gamma}\le 0.30, 
\label{eq:eq18}
\end{equation}
definitely smaller than the values used in \cite{Fr96} and \cite{Mi00,Go01}. On the other hand, the interval (\ref{eq:eq18}) is in agreement with the interval (\ref{eq:eq16}). We must stress that the value 
$g^2_{\rho\sigma\gamma}=9$ in the consideration of the process $\gamma+p\to p+\rho^0$ \cite{Fr96}, has been found for $m_{\sigma}$=0.5 GeV, which has been chosen quite arbitrarily. The $\sigma$-exchange amplitude and the corresponding estimation of 
$g_{\rho\sigma\gamma}$ depend essentially on $m_{\sigma}$. For example, in the near-threshold conditions for $\gamma+p\to p+\rho^0$, another scaling law holds for $g_{\rho\sigma\gamma}$:
\begin{equation}
\displaystyle\frac{g^2_{\rho\sigma\gamma}}{(m_{\sigma}^2+a)^2}=const,
\label{eq:eq19}
\end{equation}
where $a=M^2m_N/(m_N+M)\simeq$ 0.32 GeV$^2$ ($m_N$ is the nucleon mass). This relation shows a large correlation between $g_{\rho\sigma\gamma}$ deduced from the data about $\gamma+p\to p+\rho^0$ and the $\sigma$-mass, in such a way that larger value of $m_{\sigma}$ correspond to larger value of the coupling constant.

Moreover, the differential cross section for $\gamma+p\to p+\rho^0$, calculated in framework of $\sigma$-exchange, contains a strong dependence on the $\sigma$-mass, through the phenomenological form factor, which has to be introduced here to improve the $t-$behavior of the differential cross section $d\sigma(\gamma p\to p\rho^0)/dt$ at large $|t|$. To deduce a definite value of the $g_{\rho\sigma\gamma}$ coupling constant from a fit of the differential cross section data, it is also necessary to assume a definite value for the $g_{\sigma N N}$-coupling. Typically this value is determined from the $NN$-potential. 

The estimation of $g_{\rho\sigma\gamma}$ from the data about $\rho^0\to\pi^++\pi^-+\gamma$, which suggests the large value: $g^2_{\rho\sigma\gamma}> 10$, can not be considered very precise, as the possible $\sigma$-exchange is hidden by the large contribution of bremsstrahlung (photon radiation by charged pions). Therefore the coupling  $g_{\rho\sigma\gamma}$ derived by this decay, results in a very large width for 
$\Gamma(\rho^0\to\pi^0+\pi^0+\gamma)$, in strong contradiction with the experiment and and with the other theoretical estimations of this width. 

The HERMES-effect can not be considered a reliable source of information about the  $g_{\rho\sigma\gamma}$-coupling constant. It is more correct to say, that, in order to explain this effect in the framework of the model \cite{Mi00}, one needs large values, which are difficult to justify on the basis of what is known from the radiative decays $V\to\pi\pi\gamma$. For example, the large value of $g_{\omega\sigma\gamma}$ was justified in \cite{Mi00} by taking the lower limit for $\omega\to\pi^+\pi^-\gamma$ from \cite{PdG}, instead of a much smaller and more precise value from $\omega\to\pi^0\pi^0\gamma$. Following selection rules, the relation
$\Gamma(\omega\to\pi^+\pi^-\gamma)=2\Gamma(\omega\to\pi^0\pi^0\gamma)$ \cite{Si62} is correct independently on the decay mechanism. One deduces, for $g_{\omega\sigma\gamma}$, a value which is two orders of magnitude lower then in Ref. \cite{Mi00}.

If the small values for the $g_{\rho\sigma\gamma}$-coupling constant, as we suggest in the present paper are correct, the interpretation of the following different electromagnetic processes has to be revised:
\begin{itemize}
\item The explanation of the near-threshold cross section for $\gamma+p\to p+\rho^0$ in frame of the $\sigma$-model can not work, because the $\sigma$-contribution has to be small. The $\sigma$-contribution, taking $g^2_{\rho\sigma\gamma}\le 1$, turns out to be one order of magnitude smaller than necessary for the explanation of the existing data. In principle, one can recover by increasing correspondingly the $g_{\sigma NN}$-coupling constant, but this would seriously modify the $NN$-potential. Moreover, such increase would change the scale of the $\sigma$-contribution to the differential cross section of $\gamma+p\to p+\omega$, in the near threshold region, making the $\sigma$ and $\pi$-exchanges of the same order. So other contributions, such as Pomeron and (or) $f_2$-exchanges, which have been extrapolated up to threshold, have to be taken into account. This allows to decrease  $g_{\rho\sigma\gamma}$ up to unity \cite{La00}.
\item The explanation of the HERMES effect, in terms of coherent contribution of mesons, does not hold anymore.
\item The predictions of these constants in the framework of the $QCD$-sum rules do not hold.
\end{itemize}
\section{Acknowledgment}
We thank J. M. Laget for useful discussions on the problem of the $g_{\rho\sigma\gamma}$-coupling constant in different electromagnetic processes. One of us (M.P.R.) acknowledges the kind hospitality of Saclay, where this work was completed.
\section{Appendix}
In this Appendix we discuss the effect of the $\sigma$-width on the decay $\rho\to\pi+\pi+\gamma$.

\begin{figure}
\begin{center}
\includegraphics[height =5cm]{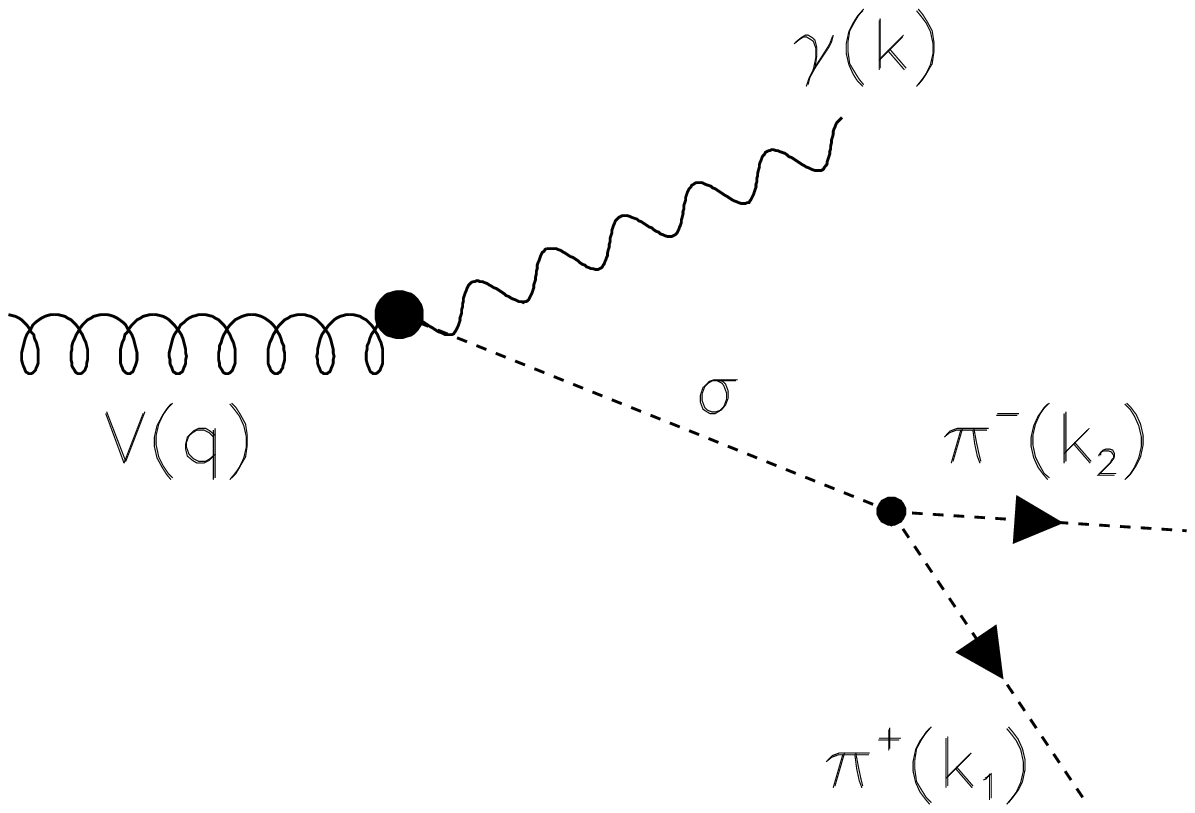}
\caption{Feynman diagrams for $V\to\pi^++\pi^-+\gamma$ decay through $\sigma$-meson production.}
\label{fig:fig4}\end{center}
\end{figure}

The matrix element, corresponding to this diagram (Fig. \ref{fig:fig4}) can be written as:
\begin{equation}
{\cal M}= {\cal M}(V\sigma^*\gamma)
\displaystyle\frac{1}{w^2-m_{\sigma}^2}m_{\sigma} g_{\sigma\pi\pi},
\label{eq:eqa1}
\end{equation}
where ${\cal M}(V\sigma^*\gamma)$ is the matrix element for the decay $V\to\sigma^*+\gamma$, with production of a virtual $\sigma$-meson, which mass $w$ (different from the mass of the 
 $\sigma$-meson, $m_{\sigma}$), coincides here with the effective mass of the produced $\pi^+\pi^-$-system, $m_{\sigma}\to m_{\sigma}-i\Gamma_{\sigma}/2$, $\Gamma_{\sigma}$ is the total width of the $\sigma$-meson.

Taking  the  expression for ${\cal M}(V\sigma^*\gamma)$, Eq.(\ref{eq:eq2}), (where the $g_{V\sigma^*\gamma}$ coupling constant does not depend, following our assumption, on the virtuality of the $\sigma$-meson), one can find, after summing over the photon polarizations and summing over the polarization of the $V-$meson:
\begin{equation}
|\overline{{\cal M}(V\sigma^*\gamma)}|^2= e^2\displaystyle\frac{g^2_{V\sigma^*\gamma}}{6}
\displaystyle\frac{(M^2-w^2)^2}{M^2}.
\label{eq:eqa2}
\end{equation}
We are interested here in the $d\Gamma/dw^2$-distribution (integrated over the effective $\pi^+\pi^-$-mass) for the decay $V^0\to\pi^++\pi^-+\gamma$ (we are considering here $\pi^+\pi^-$-production to have different final pions): 
$$
d\Gamma=
\displaystyle\frac{|\overline{\cal M}|^2}{2M}(2\pi)^4
\displaystyle\frac{d^3k}{(2\pi)^32E_{\gamma}}\int
\displaystyle\frac{d^3k_1}{(2\pi)^32E_1}
\displaystyle\frac{d^3k_2}{(2\pi)^32E_2}\delta(q-k-k_1-k_2).
$$
As $|\overline{\cal M}|^2$ depends only on the variable $w^2$, it can be taken outside the integration, so that:
\begin{equation}d\Gamma=\displaystyle\frac{|\overline{\cal M}|^2}{2^9\pi^5M}\displaystyle\frac{d^3k}{E_{\gamma}}{\cal I},
\label{eq:eqa3}
\end{equation}
where ${\cal I}$ is the result of an invariant integration (which has to be done in the CMS of the produced $\pi^+\pi^-$-system):
\begin{equation}
{\cal I}=\int \displaystyle\frac{d^3k_1}{E_1}\displaystyle\frac{d^3k_2}{E_2}\delta(q-k_1-k_2)=2\pi
\displaystyle\frac{p^*}{E^*},~E^*=\displaystyle\frac{w}{2},~
p^*=\sqrt{\displaystyle\frac{w^2}{4}-m_{\pi}^2}
\label{eq:eqa4}
\end{equation}
So ${\cal I}=2\pi\beta_w$, where $\beta_w=\displaystyle\frac{p^*}{E^*}=\sqrt{1-4 m_{\pi}^2/w^2}$ is the velocity of the pion produced in the decay of the virtual $\sigma$-meson, $\sigma^*\to\pi^++\pi^-$, with mass $w$. It is the energy of the photon, produced in the decay $V\to\pi^++\pi^- +\gamma$, which determines the effective mass of the produced $\pi^++\pi^-$-system, through the following relation:
\begin{equation}
M^2=w^2+2ME_{\gamma}
\label{eq:eqa5}
\end{equation}
which holds in the rest system of the decaying $V-$meson. Therefore we can write:
\begin{equation}
\displaystyle\frac{d^3k}{E_{\gamma}}\to 4\pi E_{\gamma}dE_{\gamma}=-\pi\left (1-\displaystyle\frac{w^2}{M^2}\right ) d w^2.
\label{eq:eqa6}
\end{equation}
Substituting Eqs. (\ref{eq:eqa4}) and (\ref{eq:eqa6}) in (\ref{eq:eqa2}) one can find:
$$
\displaystyle\frac{d\Gamma}{dw}= \displaystyle\frac{|\overline{\cal M}|^2}{2^8\pi^3M}
\left (1-\displaystyle\frac{w^2}{M^2}\right )
\sqrt{1-4 \displaystyle\frac{m_{\pi}^2}{w^2}}
$$
\begin{equation}
=\displaystyle\frac{e^2g_{V\sigma^*\gamma}}{3\pi^3 2^9} M 
\left (1-\displaystyle\frac{w^2}{M^2}\right )^3
\sqrt{1-4 \displaystyle\frac{m_{\pi}^2}{w^2}}
g^2_{\sigma^*\pi^+\pi^-}\displaystyle\frac{m_{\sigma}^2}{(w^2-m_{\sigma}^2)^2}
\label{eq:eqa7}
\end{equation}
The coupling constant $g_{\sigma\pi^+\pi^-}$ can be related to the total width $\Gamma_{\sigma}$ of the $\sigma$-meson, assuming that $\sigma\to 2\pi$ is its main decay: $\Gamma_{\sigma}=\Gamma({\sigma}\to\pi^+\pi^-)+\Gamma({\sigma}\to\pi^0\pi^0).$ In terms of the coupling constant $g=g_{\sigma\pi^+\pi^-}=g_{\sigma\pi^0\pi^0}$ (on the basis of isotopic invariance), one can write:
\begin{equation}
\Gamma({\sigma}\to\pi^+\pi^-)= \displaystyle\frac{|\overline{{\cal M}({\sigma}\to\pi^+\pi^-)}|^2}{8\pi m_{\sigma}} q= \displaystyle\frac{g^2}{16\pi} m_{\sigma}\beta_{\sigma},
\label{eq:eqa8}
\end{equation}
where $\beta_{\sigma}=\sqrt{1-4 m_{\pi}^2/m_{\sigma}^2}$ is the velocity of the pion produced in the decay of the real $\sigma$-meson, with mass $m_{\sigma}$. Taking into account the relation: $\Gamma({\sigma}\to\pi^+\pi^-)=2\Gamma({\sigma}\to\pi^0\pi^0)$, due to the identity of the two neutral pions in the decay $\sigma\to\pi^0\pi^0$, one can write:
\begin{equation}
\Gamma_{\sigma}=\displaystyle\frac{3}{2}\Gamma(\sigma\to\pi^+\pi^-)=
\displaystyle\frac{3}{32}\displaystyle\frac{g^2}{\pi} m_{\sigma}\beta_{\sigma},
\label{eq:eqa9}
\end{equation}
Substituting Eq. (\ref{eq:eqa9}) in (\ref{eq:eqa7}) one obtains:
\begin{equation}
\displaystyle\frac{d\Gamma}{dw^2}=\displaystyle\frac{e^2g_{V\sigma^*\gamma}}{144\pi^2} M 
\left (1-\displaystyle\frac{w^2}{M^2}\right )^3
\displaystyle\frac{\beta_w}{\beta_{\sigma}}\displaystyle\frac{m_{\pi}
\Gamma_{\sigma}}{(w^2-m_{\sigma}^2)^2+\Gamma_{\sigma}^2m_{\sigma}^2}
\label{eq:eqa10}
\end{equation}
So, for the full width of the decay $V\to \pi^++\pi^-+\gamma$, the following formula holds:
\begin{equation}
\Gamma(V\to \pi^+\pi^-\gamma)=\int_{4m_{\pi}^2}^{M^2}\displaystyle\frac{d\Gamma}{dw^2}dw^2.
\label{eq:eqa11}
\end{equation}
For the decay $V\to \pi^0+\pi^0+\gamma$, Eq. (\ref{eq:eqa11}) holds, divided by a coefficient 2 - due to the identity of pions in the final state.

Let us test the validity of these formulas, considering the limit of zero width, $\Gamma_{\sigma}\to 0$, i.e. considering the $ \sigma$-meson as a stable particle. We use the following symbolic relation:
$$\left |\displaystyle\frac{1}{(w^2-m_{\sigma}^2)-i\Gamma_{\sigma}m_{\sigma}}\right |^2=
\displaystyle\frac{1}{(w^2-m_{\sigma}^2)+i\Gamma_{\sigma}m_{\sigma}}~
\displaystyle\frac{1}{(w^2-m_{\sigma}^2)-i\Gamma_{\sigma}m_{\sigma}} \stackrel{\Gamma_{\sigma}\to 0}{\rightarrow} $$
$$=\displaystyle\frac{1}{(w^2-m_{\sigma}^2)+i\Gamma_{\sigma}m_{\sigma}}
\left [ \displaystyle\frac{1}{w^2-m_{\sigma}^2}+i\pi\delta(w^2-m_{\sigma}^2)\right ]\to
\displaystyle\frac{\pi}{m_{\sigma}\Gamma_{\sigma}}\delta(w^2-m_{\sigma}^2)
$$
This $\delta$-function allows one to integrated easily Eq. (\ref{eq:eqa11}), getting the following result:
\begin{equation}
\Gamma(V\to \sigma\gamma\to \pi^+\pi^-\gamma)=\displaystyle\frac{\alpha}{36}g^2_{V\sigma\gamma}M
\left (1-\displaystyle\frac{m_{\sigma}^2}{M^2}\right )^3.
\label{eq:eqa12}
\end{equation}
To find the radiative width for the radiative decay $V\to\sigma\gamma$, taking into account both possibilities, $\sigma\to\pi^++\pi^-$ and $\sigma\to\pi^0+\pi^0$, we have to introduce a factor 3/2, so that:
$$
\Gamma(V\to \sigma\gamma)=\displaystyle\frac{\alpha}{24}g^2_{V\sigma\gamma}M
\left (1-\displaystyle\frac{m_{\sigma}^2}{M^2}\right )^3
$$
in agreement with the direct calculation, see 
Eq. (\ref{eq:eq3}).

\end{document}